\documentclass[a4paper,11pt]{article}
\pdfoutput=1 

\usepackage{jinstpub} 
\usepackage{float} 
\usepackage{graphicx} 
\usepackage{subfigure}
\usepackage{wrapfig}
\usepackage{siunitx}
\DeclareSIUnit\parsec{pc}
\DeclareSIUnit\lightyear{ly}
\DeclareSIUnit\DAY{day}
\DeclareSIUnit\YEAR{year}
\DeclareSIUnit\speedOfLight{c}
\usepackage{pgf,tikz,pgfplots}
\usepackage{url}

\title{A resistive ACHINOS multi-anode structure with DLC coating for spherical proportional counters}

\author[a]{I.~Giomataris,}
\author[a]{M.~Gros,}
\author[b]{I.~Katsioulas,}
\author[a,b,1]{P.~Knights\note{Corresponding author.},}
\author[a]{J.-P.~Mols,}
\author[b]{T.~Neep,}
\author[b]{K.~Nikolopoulos,}
\author[c]{G.~Savvidis,}
\author[d]{I.~Savvidis,}
\author[e]{L.~Shang,}
\author[b]{R.~Ward}
\author[f]{and Y.~Zhou}

\affiliation[a]{IRFU, CEA, Universit\'e Paris-Saclay,\\ Gif-sur-Yvette F-91191, France}
\affiliation[b]{School of Physics and Astronomy, University of Birmingham,\\Birmingham B15 2TT, United Kingdom}
\affiliation[c]{Department of Physics, Queen's University,\\ Kingston, ON K7L 3N6, Canada}
\affiliation[d]{Aristotle University of Thessaloniki,\\ Thessaloniki, Greece}
\affiliation[e]{State Key Laboratory of Solid Lubrication, Lanzhou Institute of Chemical Physics, \\Chinese Academy of Science, Lanzhou 730000, China}
\affiliation[f]{State Key Laboratory of Particle Detection and Electronics, University of Science and Technology of China,\\ Hefei 230026, China}

\emailAdd{prk313@bham.ac.uk}

\abstract{The spherical proportional counter is a gaseous detector
  used in a variety of applications, including direct dark matter 
  and neutrino-less double beta decay searches. The ACHINOS multi-anode structure is a read-out technology 
  that overcomes the limitations of single-anode read-out structures for large-size detectors 
  and operation under high pressure.
  A resistive ACHINOS is presented, where the 3D printed central component is coated in a Diamond-Like Carbon (DLC) layer.
  The production and testing of the
  structure, in terms of stability and resolution, is described.
}

\keywords{Dark Matter detectors (WIMPs, axions, etc.); Detector design and construction technologies and materials; Electronic detector readout concepts (gas, liquid); Gaseous detectors}

\begin{document}
\maketitle
\flushbottom

\section{Introduction}
\begin{figure}[!h]
\centering
\subfigure[\label{fig:sphereScheme}]{\includegraphics[width=0.45\textwidth]{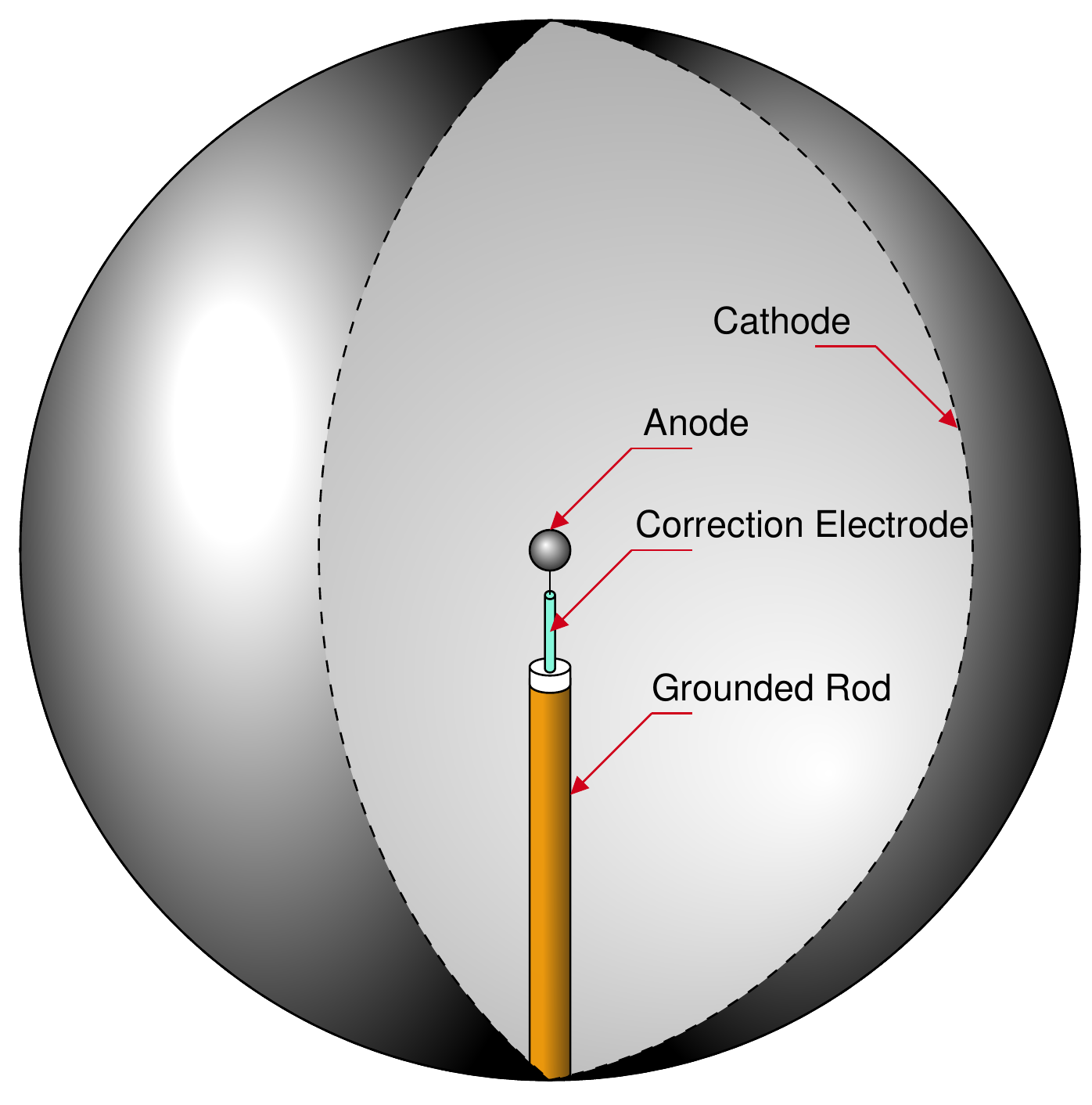}}
\subfigure[\label{fig:achinosBasicScheme}]{\includegraphics[width=0.4\textwidth]{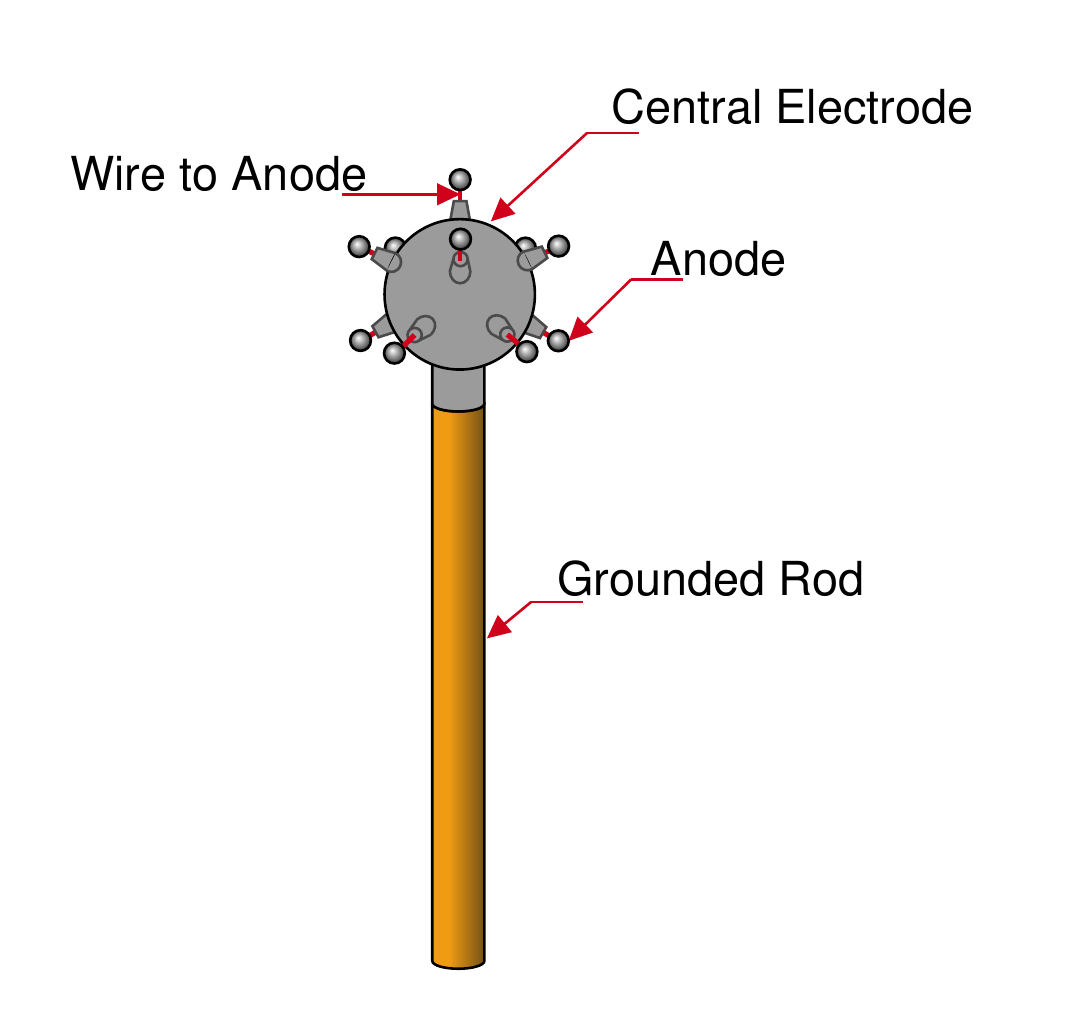}}
\caption{
Schematic of \subref{fig:sphereScheme} the spherical proportional counter, showing the grounded cathode, central read-out anode, and the anode support structure, and \subref{fig:achinosBasicScheme} the ACHINOS structure.
}
\end{figure}

In its simplest form, the spherical proportional
counter~\cite{Giomataris2008-mx}, shown in figure~\ref{fig:sphereScheme},
 consists of a grounded, spherical, metallic
vessel filled with an appropriate gas mixture and a spherical anode of radius approximately 
$1$\;\si{\milli\meter} at the centre. The anode is supported by a 
grounded metallic rod, which also shields the
wire 
used to apply a positive voltage to the anode and read
out the signal. 
The electric field, which varies as $1/r^{2}$ in an ideal spherical proportional counter, allows the
ionisation electrons produced through particle interactions in the gas volume to
drift to the anode.
Within approximately $1\;\si{\milli\meter}$ from the anode an
avalanche occurs, providing signal amplification.
Experiments using this detector include
NEWS-G~\cite{Arnaud2018-nr}, searching for dark matter
particles, and R2D2~\cite{Meregaglia_2018}, searching for
neutrino-less double beta decay.

For large detectors, or detectors operating in high pressure, the small ratio of electric field strength to gas pressure (E/P) increases the probability of electron attachment and recombination. 
In the single anode configuration, increasing the E/P ratio can be achieved by either increasing the anode voltage or the anode size, both of these options lead to an increased discharge probability. 

ACHINOS, a sensor structure composed of several anodes at a radius
$r_S$ from the centre, as shown in
figure~\ref{fig:achinosBasicScheme}, has been proposed to overcome
this challenge~\cite{Giganon2017-jinst}. The anodes are maintained in
position by means of a central support structure.  This structure
needs to be constructed using resistive materials, as has been the
case with previous read-out technologies~\cite{Katsioulas:2018pyhSUB}.
With ACHINOS, the avalanche gain is determined by the anode radius
$r_A$ and voltage $V$, while the electric field at large radii is the
collective electric field of all the anodes, determined by $V$, $r_S$,
and the number of anodes.  Additionally, it is possible to read out
each anode individually, allowing the three-dimensional reconstruction
of the ionisation tracks.

\section{ACHINOS with ``resistive glue'' coating}
The ACHINOS central electrode is manufactured with resistive material,
a choice that has been shown to reduce sparking rate and
intensity~\cite{Giganon2017-jinst,Katsioulas:2018pyhSUB}.
\begin{figure}[!b]
\centering
\includegraphics[width=0.45\linewidth]{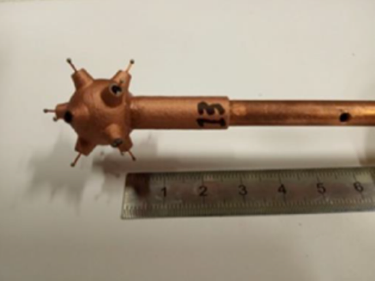}
\caption{ACHINOS using an Araldite-copper layer on the 3-D-printed structure to form the central electrode.\label{fig:cuAchinos}}
\end{figure}
3D printing provides a convenient means to produce a high-precision structure, however, currently 3D printing with insulators and conductors is more widely available.
Thus, an appropriate coating needs to be applied. 
Initially, an Araldite adhesive mixed with copper powder was explored, with main benefits being the relatively low radioactivity of Araldite 2011~\cite{BUSTO200235} 
and the possibility to control the resistivity of the coating by changing the relative amounts of Araldite and copper.
It was found that mixtures containing $20\%$ to $50\%$ w/w copper powder were appropriate.
An example is shown in figure~\ref{fig:cuAchinos}. 
Despite the promising results, 
the coating layer was found to be susceptible to damage from
discharges. 
Specifically, a single discharge could destroy the central
electrode coating, by creating a conductive path. This behaviour was also
observed repeatedly using a spark-test chamber.

\section{ACHINOS using DLC coating}
DLC is a form of amorphous carbon containing both the
diamond and the graphite crystalline phase. 
Thanks to its excellent surface resistivity, in addition to
structural, chemical and thermal stability, DLC
coating~\cite{LV2020162759} offers a novel method for producing 
high quality resistive materials for gaseous
detectors~\cite{ZHOU201931,ATTIE2020163286}. 

\begin{figure}[!ht]
\centering
\subfigure[\label{fig:dlc1}]{\includegraphics[width=0.6\textwidth]{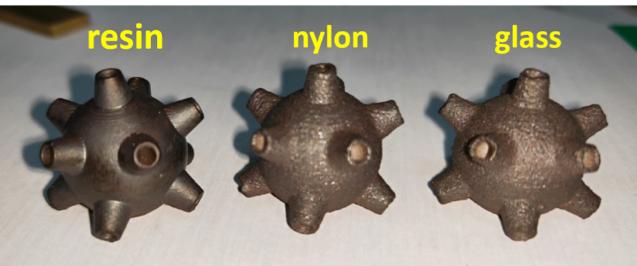}}
\subfigure[\label{fig:achinos3}]{\includegraphics[width=0.2465\textwidth, angle=90]{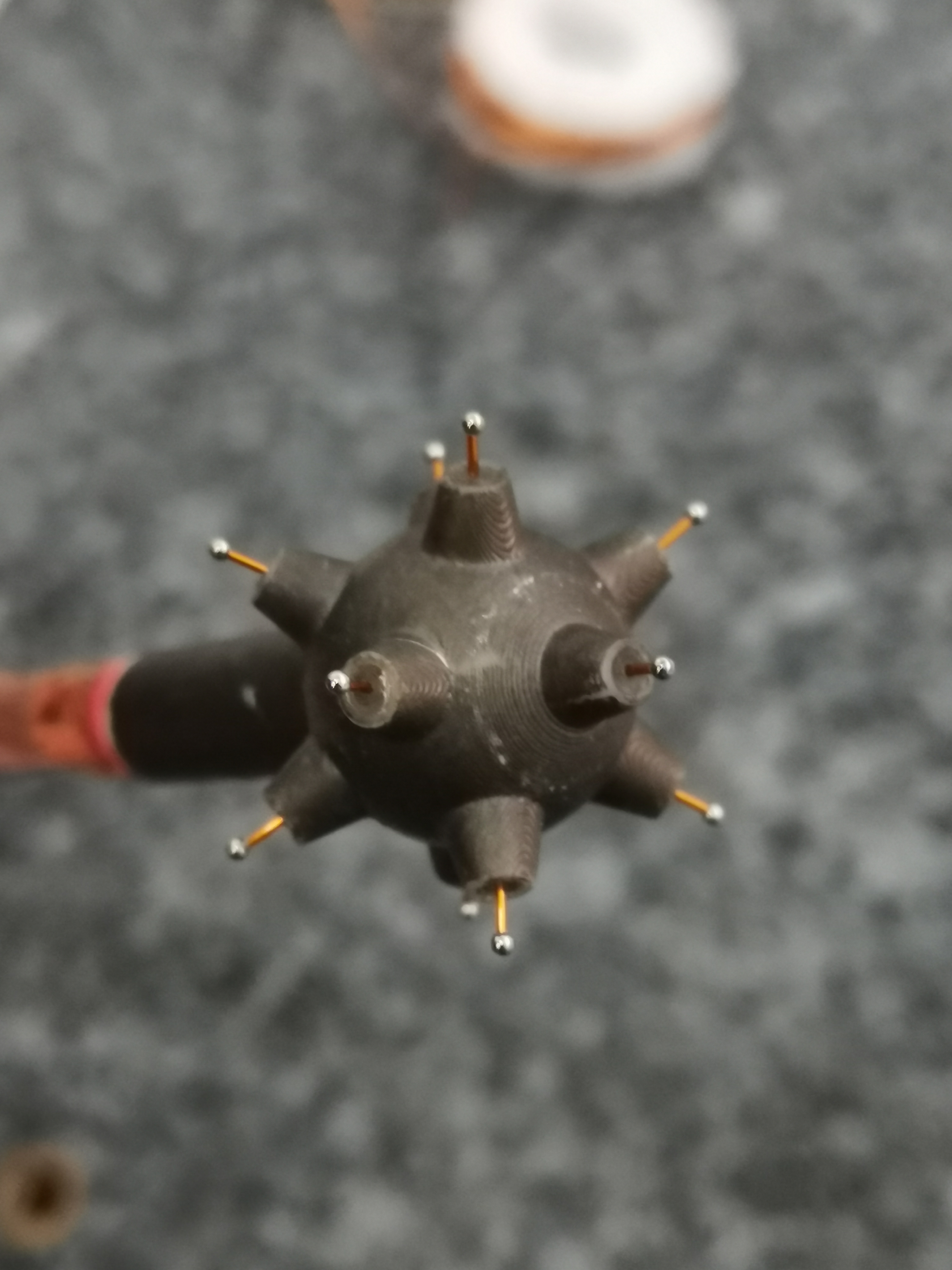}}
\caption{\subref{fig:dlc1} Three different support materials (resin, nylon, glass) covered with a DLC layer. \subref{fig:achinos3} An 11-anode ACHINOS constructed using a DLC-coated support structure.  
\label{fig:DLC}}
\end{figure}

The central structure was constructed using 3D printing with different substrates, including
resin, nylon, and
glass, as shown in figure~\ref{fig:dlc1}. 
DLC was deposited using magnetron sputtering.
Several batches of DLC-coated structures were produced, with a range
of coating thickness between $360\;\si{\nano\meter}$ and
$720\;\si{\nano\meter}$. The measured resistance between two
anti-diametric points on the surface, ranged from
$0.3$ to $10\;\si{\giga\ohm}$.
The DLC-coated resin 3D printed structure was mounted on a copper rod and electrically connected to it using a conductive Araldite copper mixture, as shown in figure~\ref{fig:achinos3}. 

\section{Experimental results}
The ACHINOS structure with 11-anodes, each $1\;\si{\milli\meter}$ in diameter, shown in figure~\ref{fig:achinos3} was installed in a $30\;\si{\centi\meter}$ diameter spherical proportional counter that operated in sealed mode. An ${}^{55}$Fe source was installed inside the detector, the position of which could be adjusted without opening the detector. The source principally emits a $5.9\;\si{\kilo\eV}$ X-ray when it decays to ${}^{55}$Mn by electron capture. The experimental set-up is shown in figure~\ref{fig:experimentalScheme}. 
An ISEG NHR 22 60r power supply was used to apply voltage to the anodes. The central electrode was, typically, grounded through the rod, but alternative configurations were also tested. 
Signals are read out by a CREMAT
CR-110-R2 charge sensitive preamplifier. The
preamplifier output is passed to 
a ``CALI'' digitiser, manufactured by CEA Saclay~\cite{adfthesis}, with a dynamic range of
$\pm 1.25\;\si{\volt}$ and maximum sampling frequency of $5\;\si{\mega\Hz}$.

\begin{figure}[!t]
\centering
\subfigure[\label{fig:experimentalScheme}]{\includegraphics[width=0.45\textwidth]{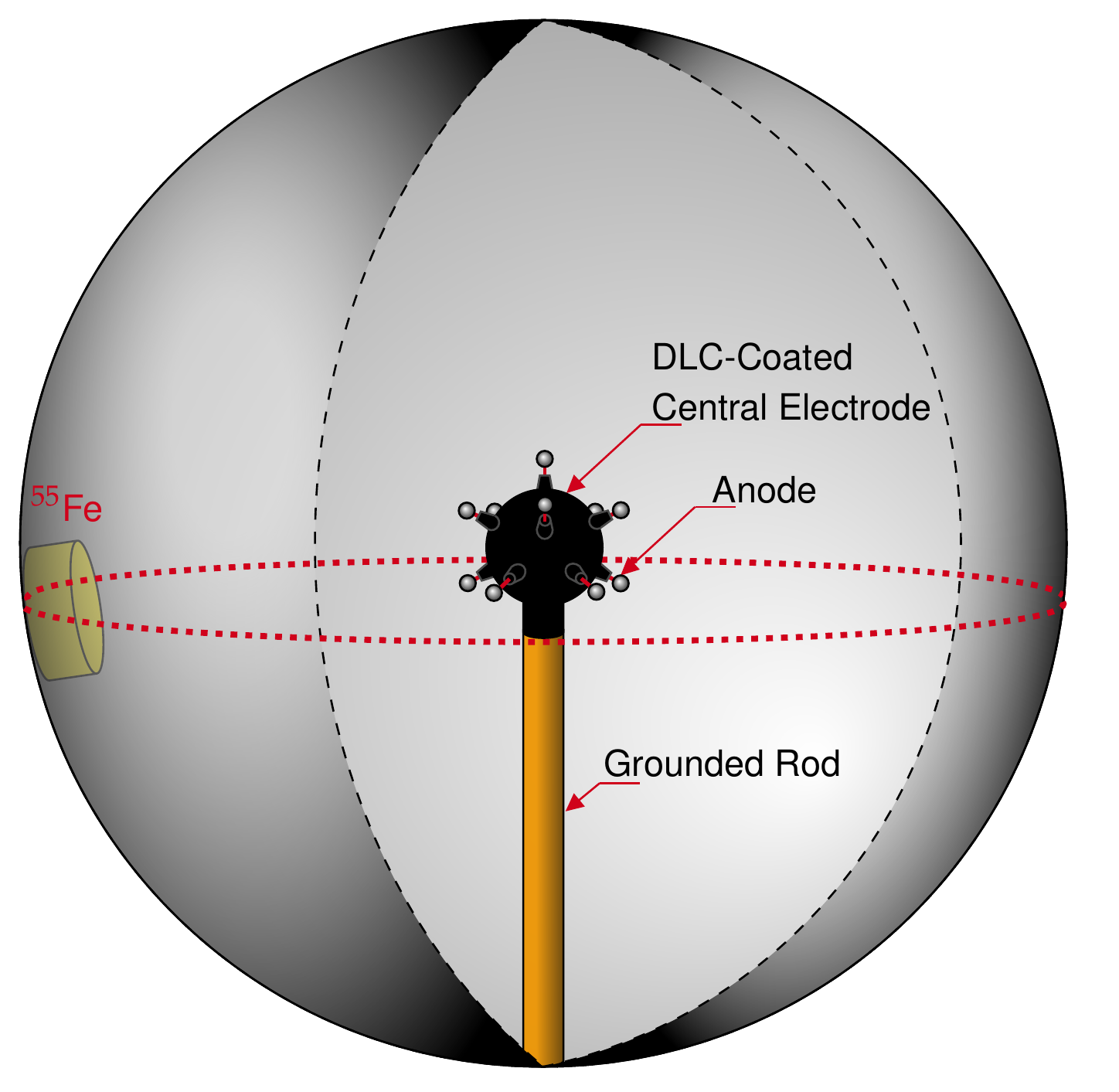}}
\subfigure[\label{fig:examplePulse200}]{\includegraphics[width=0.49\textwidth]{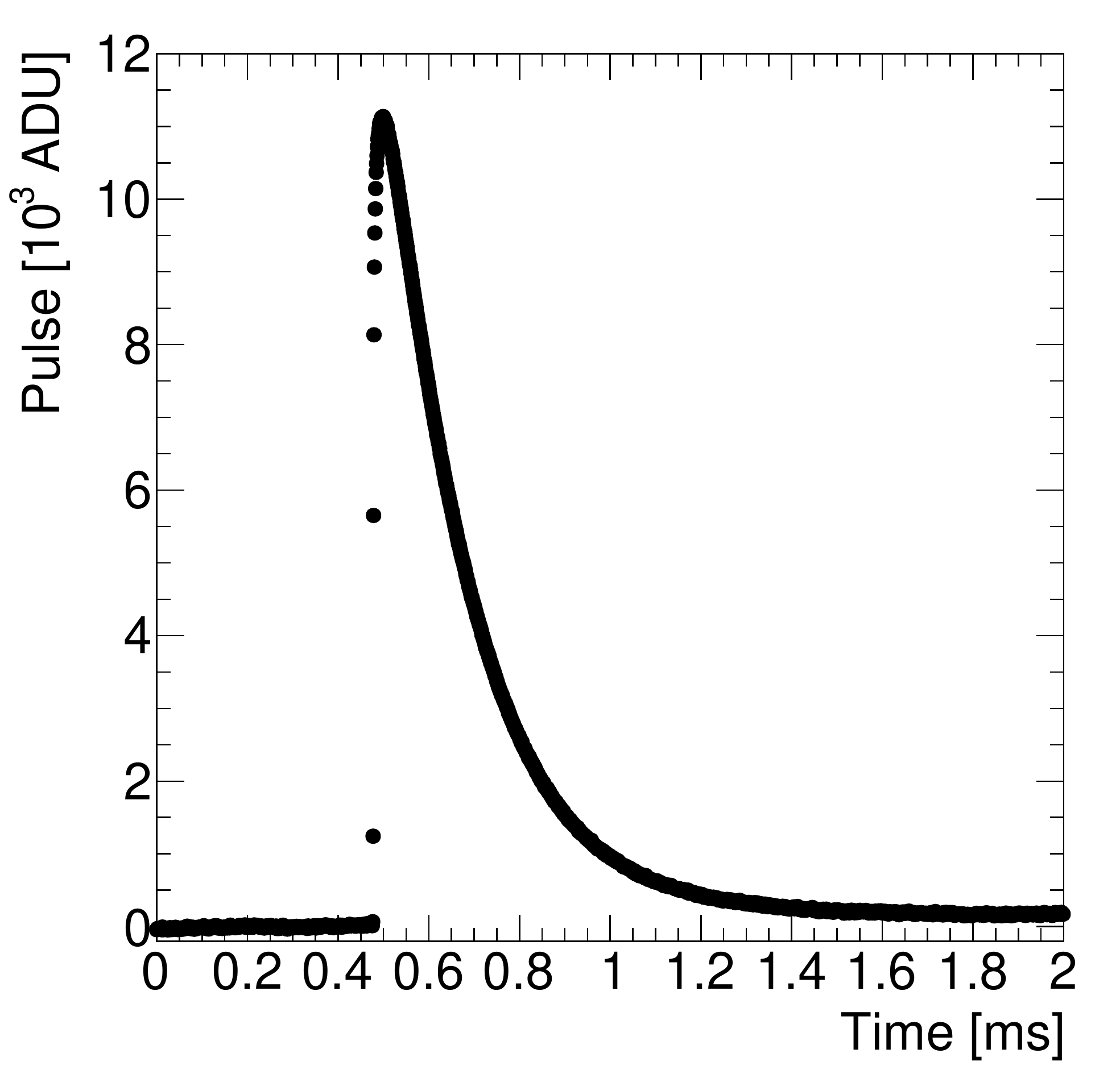}}
\caption{\subref{fig:experimentalScheme} Schematic of the experimental
  set-up, with the position of the ${}^{55}$Fe source relative to the 11-anode
  ACHINOS\label{fig:experimentalScheme}. \subref{fig:examplePulse200}
  Digitised output pulse recorded in a $2\;\si{\milli\second}$ window,
  with the pulse peak positioned at $25\%$ of the window width. 
\label{fig:DLC}}
\end{figure}

An example digitised output pulse, recorded with a sampling
rate of $1\;\si{\mega\Hz}$, is shown in figure~\ref{fig:examplePulse200}.
The exponential falling edge is defined by the
$140\;\si{\micro\second}$ time constant of the preamplifier.
A preamplifier output of $1\;\si{ADU}$,\footnote{ADU = Analogue-to-Digital Units.} corresponds to a charge of
approximately $0.027\;\si{\femto \coulomb}$ at the preamplifier input. 
Contamination from cosmic-ray muons interacting in the gas volume is
suppressed using loose event selections on the pulse rise time,
defined as the time required for the signal to increase from $10\%$ to
$90\%$ of its amplitude, and the ratio of the pulse integral to its amplitude.

\subsection{Gain}
\begin{figure}[h!]
\centering
\includegraphics[width=0.49\linewidth]{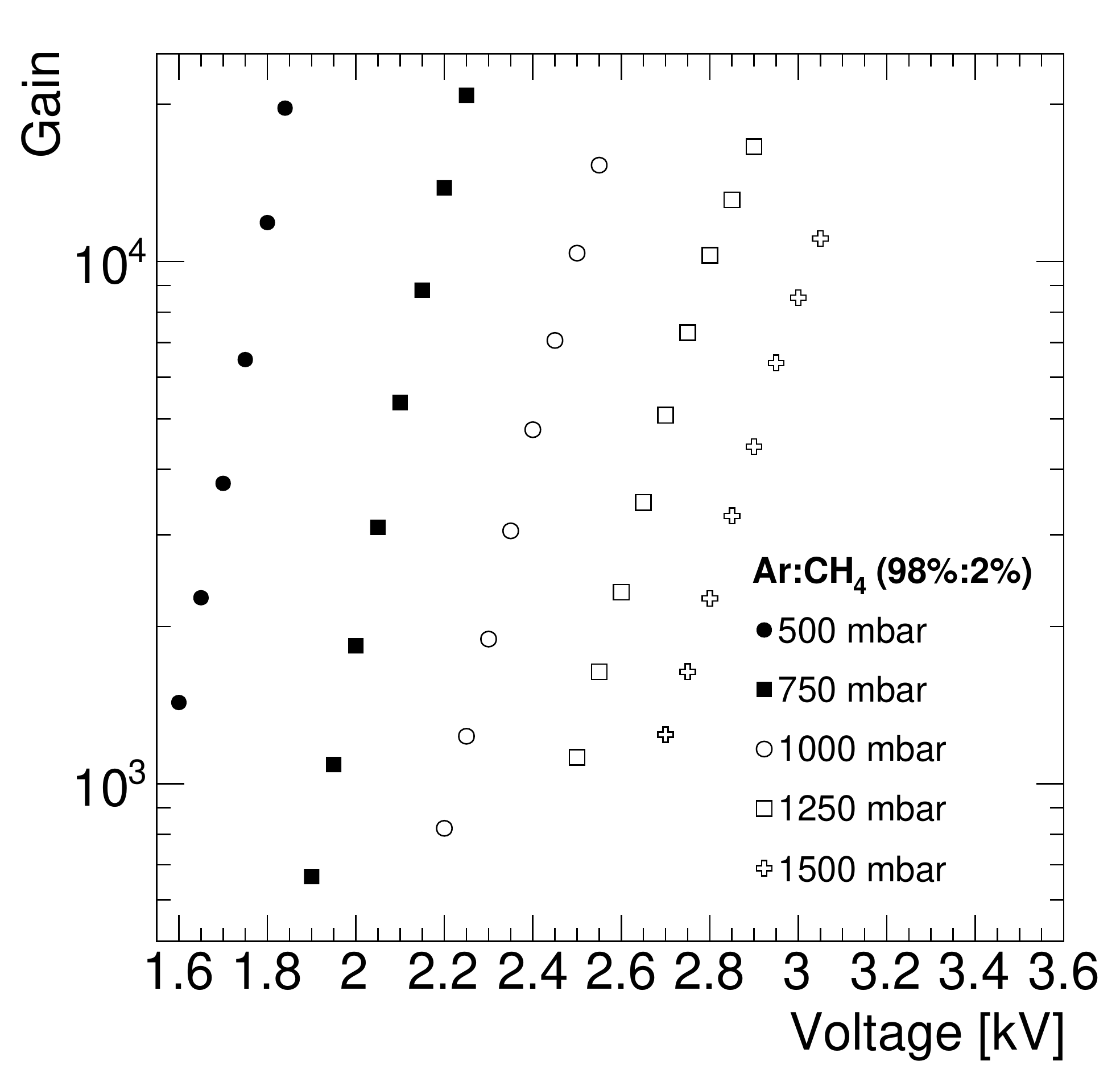}
\caption{Measured amplitude versus the voltage applied to the anode for several pressures of Ar$\,:\,$CH${4}$ ($98\%\,$:$\,2\%$) in a spherical proportional counter using an ACHINOS. 
\label{fig:gainStudy}}
\end{figure}

The detector was filled with $500\;\si{\milli\bar}$ of
Ar$\,:\,$CH${4}$ ($98\%\,$:$\,2\%$) and data were collected at various anode
voltages and pressures, and the amplitude of the
$5.9\;\si{\kilo\eV}$ peak were recorded. These measurements, performed with the 
central electrode electrically floating, are presented in
figure~\ref{fig:gainStudy}, where the detector is shown to operate in proportional mode
over a wide pressure range. Furthermore it is
demonstrated that large gas gains (avalanche electron multiplication factor) of up to $\num[retain-unity-mantissa
= false]{1e4}$ can be achieved with this detector configuration.

\subsection{Anode response uniformity}
\begin{figure}[h!]
\centering
\includegraphics[width=0.49\linewidth]{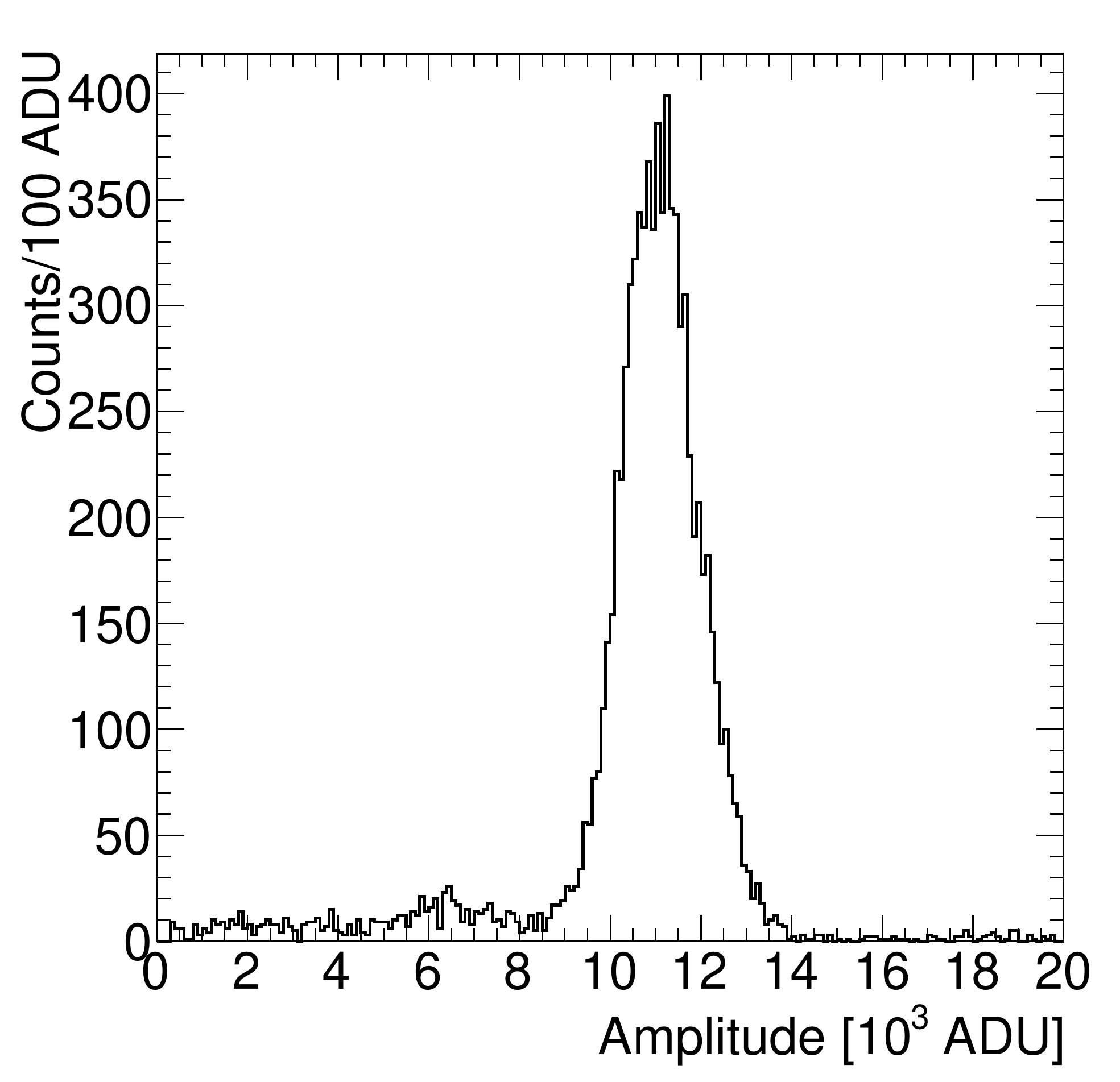}
\caption{
Energy spectrum from an ${}^{55}$Fe source measured using a spherical proportional counter filled with $1000\;\si{\milli\bar}$ of Ar$\,:\,$CH${4}$ ($98\%\,$:$\,2\%$)  and using an ACHINOS. The primary peak has an energy resolution ($\sigma$) of $(7.4\pm0.1)\%$. The second peak, to the left of the main one, is the argon escape peak. 
\label{fig:ua16e008_amp_cut}
}
\end{figure}

\begin{figure}[!h]
\centering
\subfigure[\label{fig:anglePlot}]{\includegraphics[width=0.49\textwidth]{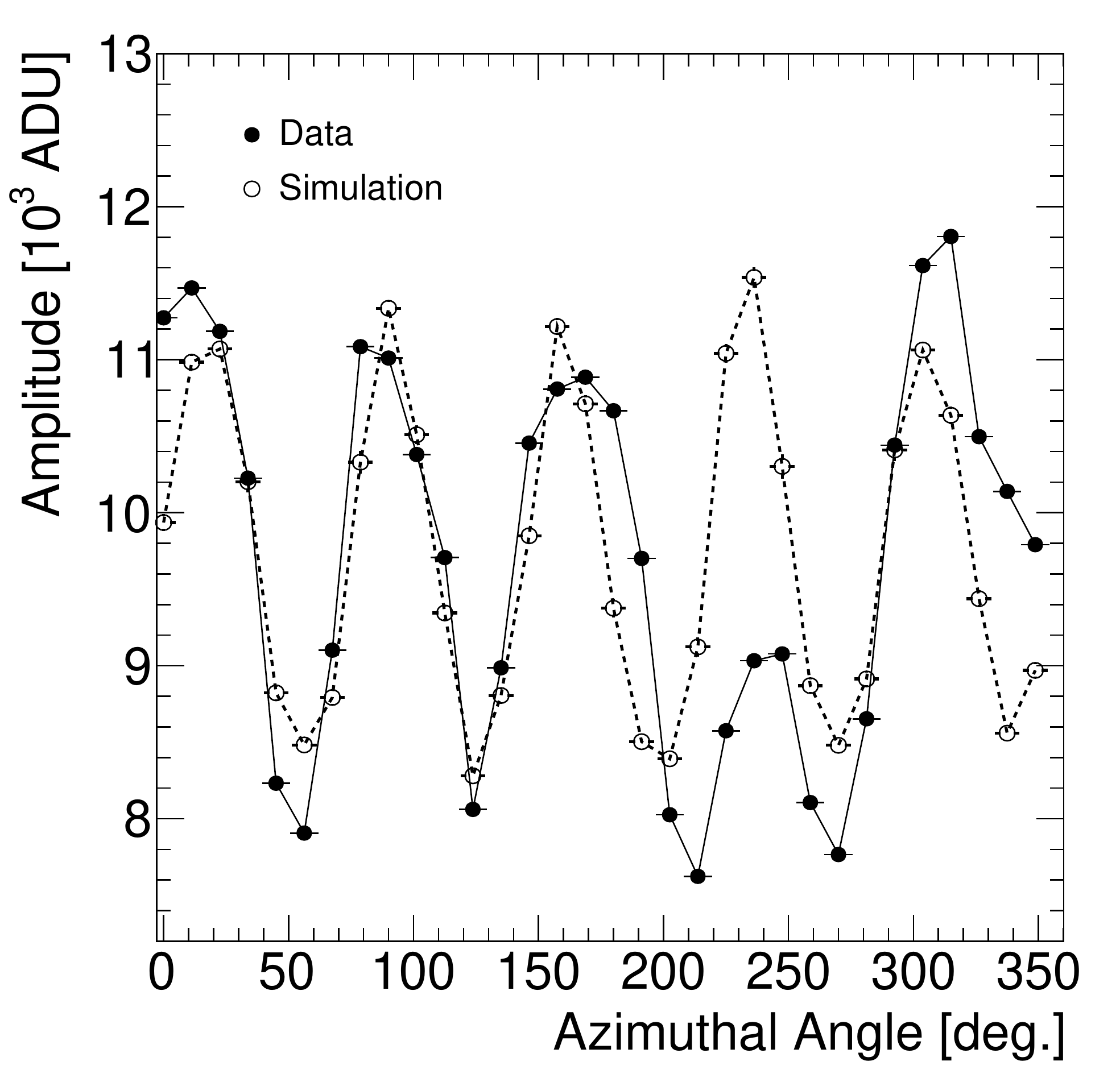}}
\subfigure[\label{fig:angleResPlot}]{\includegraphics[width=0.49\textwidth]{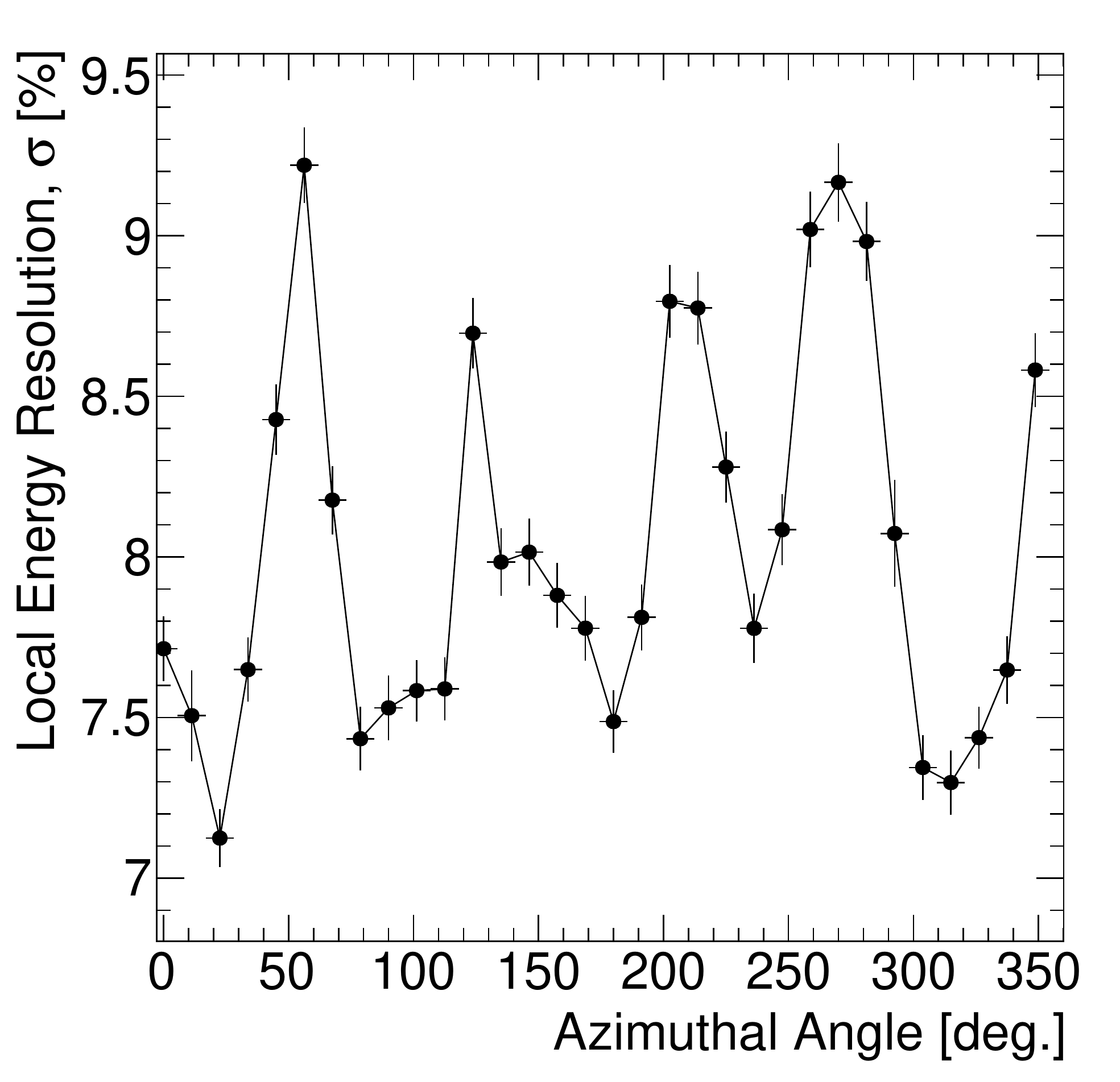}}
\caption{Measured and simulated amplitude \subref{fig:anglePlot} and measured local energy resolution $\sigma$ \subref{fig:angleResPlot} of the $5.9\;\si{\kilo\eV}$ X-ray in a spherical proportional counter with an ACHINOS as a function of azimuthal angle. The detector was operated filled with $1000\;\si{\milli\bar}$ of Ar$\,:\,$CH${4}$ ($98\%\,$:$\,2\%$). }
\end{figure}

To check the uniformity of the sensor response, data were taken with the ${}^{55}$Fe
source at various longitudinal positions, while at a fixed latitude of approximately
$12^{\circ}$ below the equator, as shown schematically in
figure~\ref{fig:experimentalScheme}.
The detector was operated with $1000\;\si{\milli\bar}$ of Ar$\,:\,$CH${4}$ ($98\%\,$:$\,2\%$)  and an anode voltage of 
$2200\;\si{\volt}$. 
The amplitude and local energy resolution were estimated using the $5.9\;\si{\kilo\eV}$ peak and an example spectrum is shown in figure~\ref{fig:ua16e008_amp_cut}. A gaussian fit of this yielded an amplitude of $(11090\pm10)\;\si{ADU}$, corresponding to a gas gain of $\num{8.3E3}$, and a local energy resolution $\sigma$ of $(7.4\pm0.1)\%$. The same procedure was repeated for the other measurements, with the results shown in figure~\ref{fig:anglePlot} for the amplitude and figure~\ref{fig:angleResPlot} for the local energy resolution, which varies between $7.1\%$ and $9.2\%$.

The maximum (minimum) amplitude at approximately $320^{\circ}$ ($210^{\circ}$) corresponds to a gas gain of $\num{8.8E3}$ ($\num{5.7E3}$).
The amplitude modulation as a function of the azimuthal angle observed in figure~\ref{fig:anglePlot} is intriguing. To understand its origin a dedicated framework for the simulation of spherical proportional counters was used~\cite{Katsioulas:2020ycw}, which combines the strengths of Geant4~\cite{Allison:2016lfl} and Garfield++~\cite{Veenhof:1998tt}. In the simulation 5.9 keV photons were generated near the cathode surface and directed within a $45^{\circ}$ cone towards the detector centre. A similar analysis to those applied to the data was implemented. The open circles in figure~\ref{fig:anglePlot} correspond to the simulation results which reproduce the modulation observed in the data. 

\begin{figure}[!h]
\centering
\subfigure[\label{fig:simOverlaid}]{\includegraphics[width=0.49\textwidth]{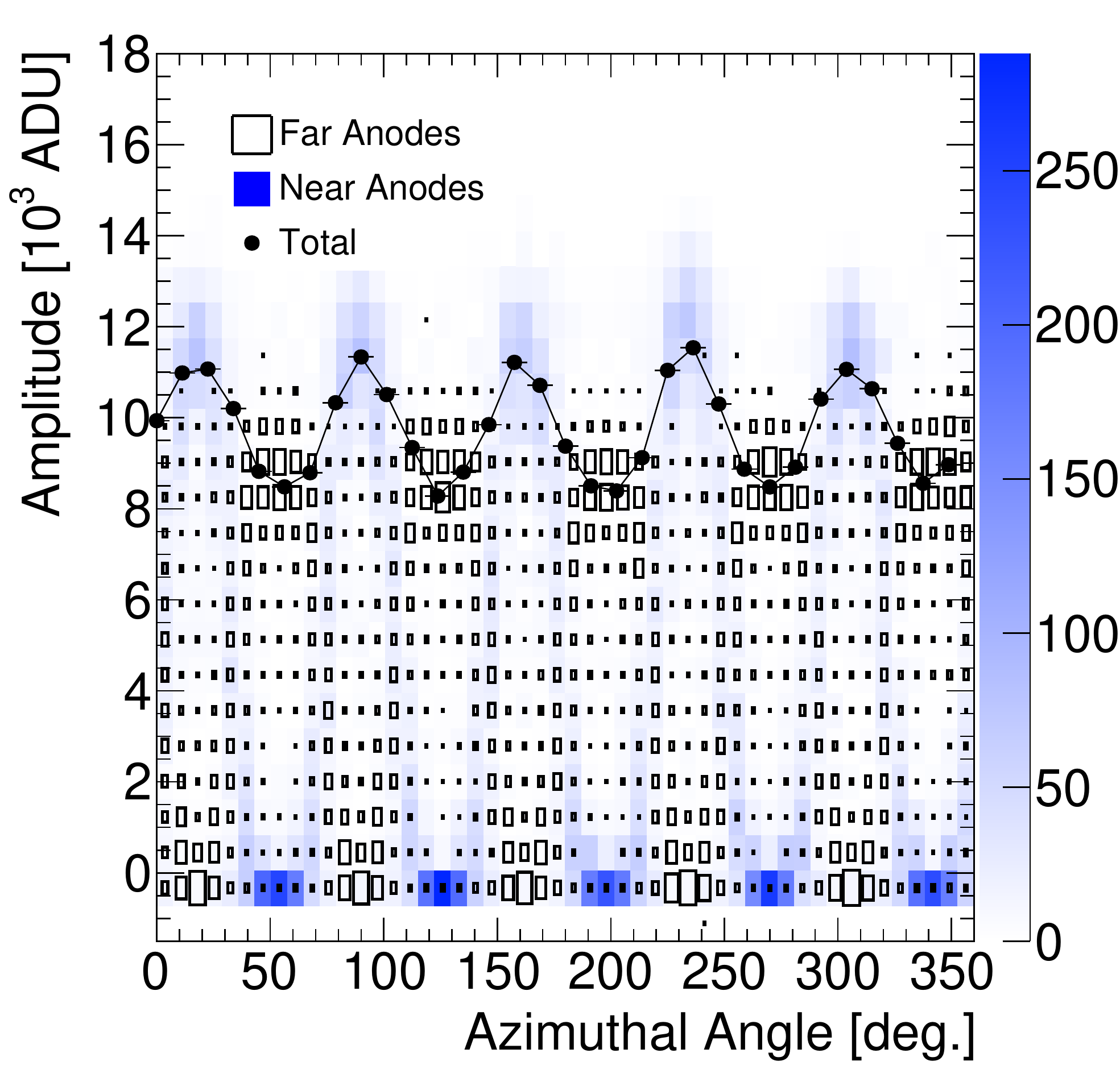}}
\subfigure[\label{fig:simOverlaidDiffV}]{\includegraphics[width=0.49\textwidth]{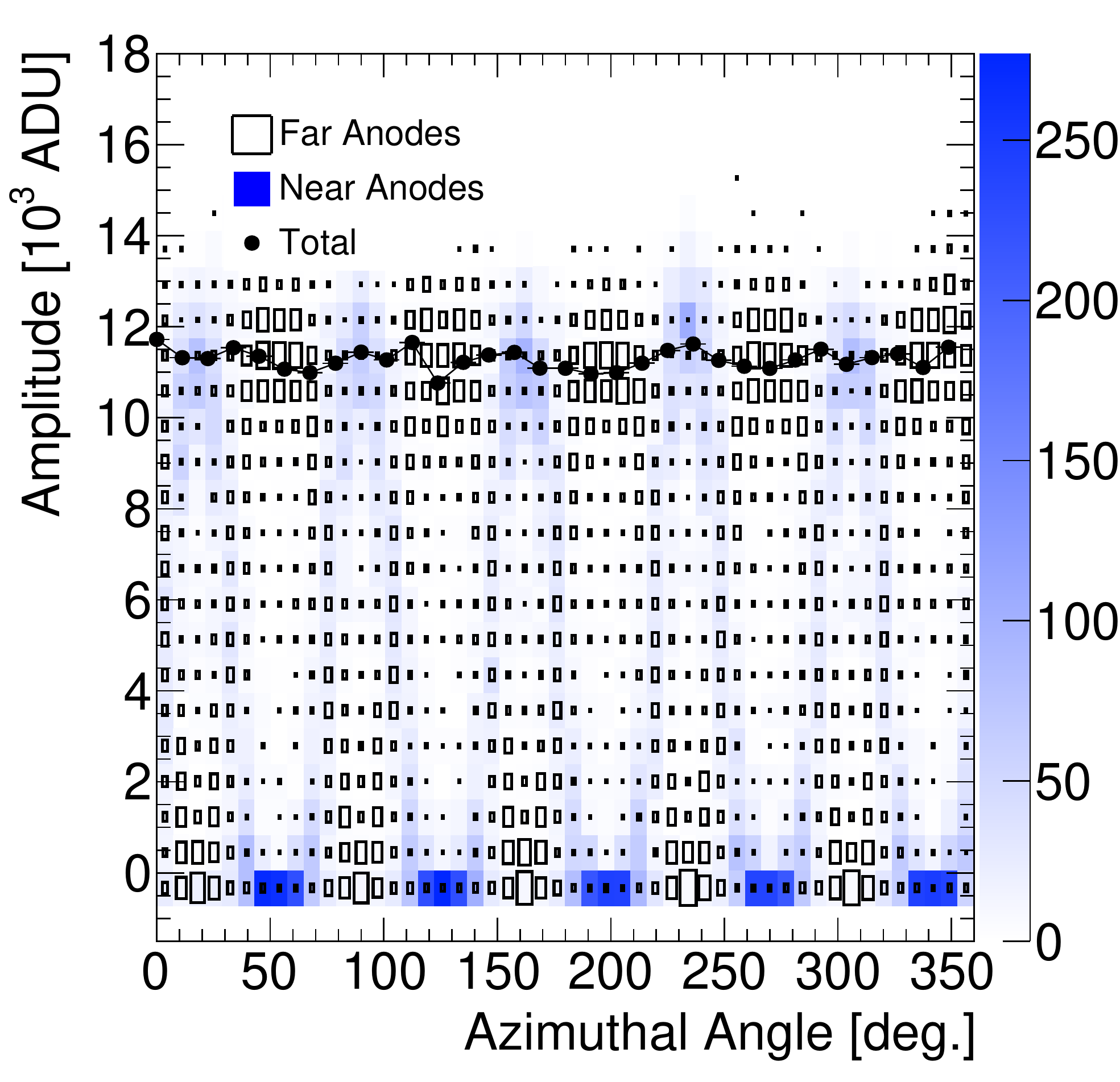}}
\caption{\subref{fig:simOverlaid} Simulated amplitude recorded by the Near and Far anodes as a function of azimuthal angle.  The difference in relative maximum amplitude between the two is due to a higher electric field magnitude for the Near anodes, which is caused by their proximity to the rod. \subref{fig:simOverlaidDiffV} The amplitude recorded by the Near and Far anodes in the case where $30\;\si{\volt}$ more is applied to the Far anodes. }
\end{figure}

To understand the origin of the effect, the 11-anodes of the ACHINOS
were split in two groups. The first group comprises the five co-planar
anodes near the support rod, while the second group comprises the five
co-planar anodes far from the support rod and the anode furthest away
from the support rod. In the following these two groups will be
referred to as ``Near Anodes'' and ``Far Anodes'' respectively. In
figure~\ref{fig:simOverlaid} the amplitudes recorded by reading out
each set separately is presented. It is observed that the amplitudes
are modulating in antiphase. This provides a clear explanation of the
effect; as the source is moved around the sphere, the produced primary
electrons are drifting towards different anodes, alternating between
the Near and the Far anodes.

The relative difference in maximum amplitude between the Near and Far
anodes is due to the gain difference between the two sides. The Near
anodes exhibit an increased electric field due to the proximity to the
grounded support rod, which results in deviations from the spherical
symmetry. This effect can be readily corrected for by separately
adjusting the applied voltage to the Near and Far anodes. In
figure~\ref{fig:simOverlaidDiffV} the simulation is repeated with the
voltage applied to the Far anodes increased by $30\;\si{\volt}$. In
this case the azimuthal dependence of the amplitude is minimised, and
could be eliminated by further tuning of the Far anodes
voltage.

A potential source of additional response non-uniformity may arise
from construction differences of the individual anodes. This effect is
not included in the simulation, however it can be estimated from the
experimental data. The difference in gain within the two sets of
co-planar anodes are typically within $10\%$. This is shown in figure~\ref{fig:anglePlot}, where a single outlier
is also observed in each of the sets. These differences are attributed to
variations in the anode radii, the smoothness of the anode surface,
and the distance to the central electrode. These non uniformities can be
corrected for with an anode-by-anode calibration, when each anode is
read-out individually.

\subsection{Stability}
\begin{figure}[h]
\centering
\includegraphics[width=0.49\linewidth]{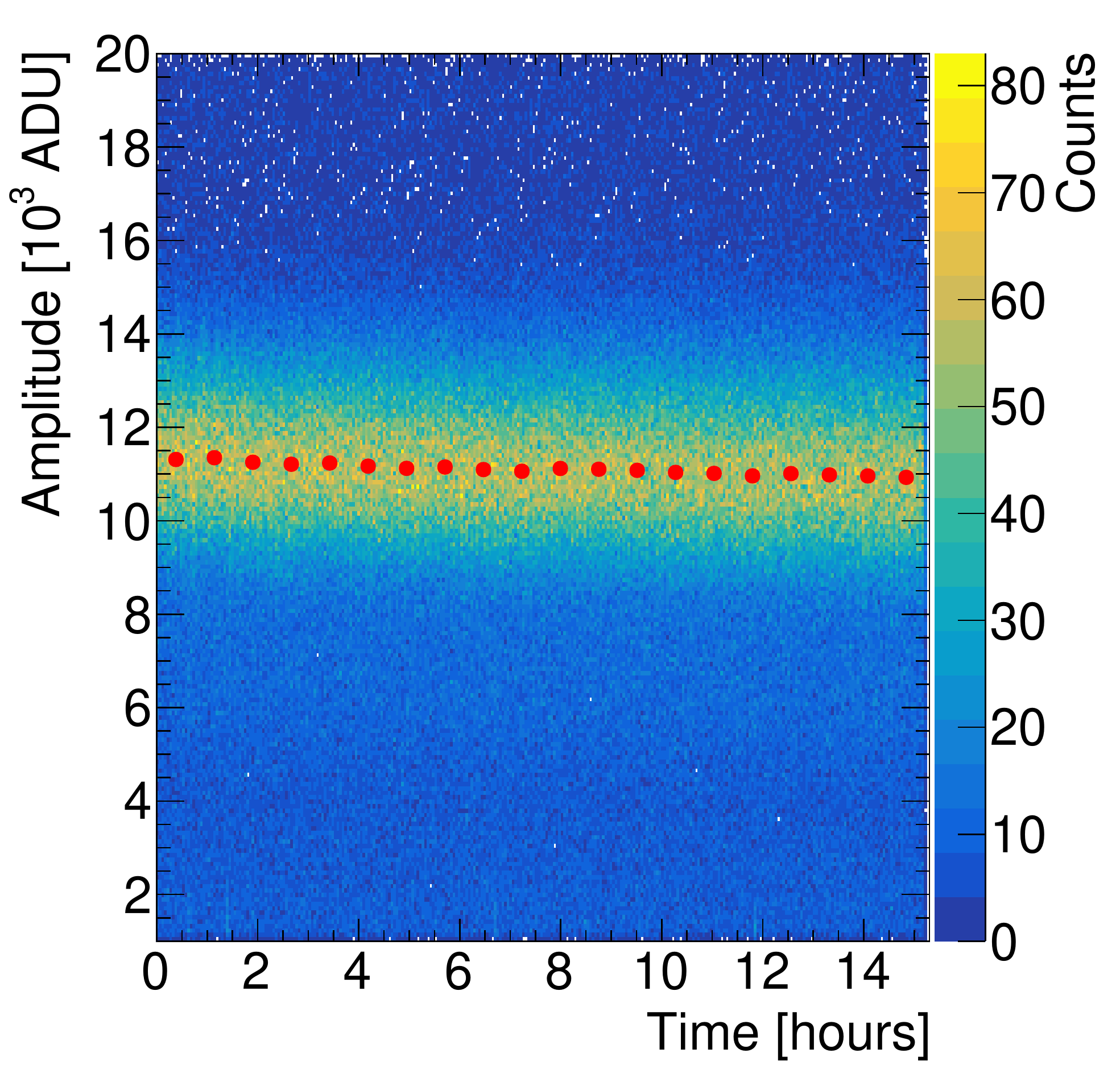}
\caption{
Amplitude versus time for a spherical proportional counter filled with $1000\;\si{\milli\bar}$ of Ar$\,:\,$CH${4}$ ($98\%\,$:$\,2\%$) and using an ACHINOS. The red points superimposed on the histogram show the mean amplitude in time slices. The slight decrease in amplitude with time is attributed to impurities leaking into the detector. 
\label{fig:stability}
}
\end{figure}

The detector operated stably filled with $1000\;\si{\milli\bar}$ of
Ar$\,:\,$CH${4}$ ($98\%\,$:$\,2\%$)  and the anode voltage set to
$2200\;\si{\volt}$.  The amplitude of the ${}^{55}$Fe peak was
monitored and no significant gain variations were observed over
approximately $15\;\si{hours}$, as shown in figure~\ref{fig:stability}. The gradual decrease in gain over time
is attributed to the introduction of contaminants coming from
outgassing and leaks in the vacuum system, a behaviour that has been
observed in earlier studies~\cite{Katsioulas:2018pyhSUB}.

In further tests, the sensor was operated without issues for a period
of 30 days despite an observed average of approximately one discharge per
day. Furthermore, no damage was observed following 
intentionally induced discharges.

\section{Future developments}
Additional R\&D effort is foreseen, aiming to consolidate the
construction and performance of ACHINOS. The main directions for such developments are
briefly discussed in the following.

\begin{figure}[!h]
\centering
\subfigure[\label{fig:bhamjig}]{\includegraphics[width=0.445\textwidth]{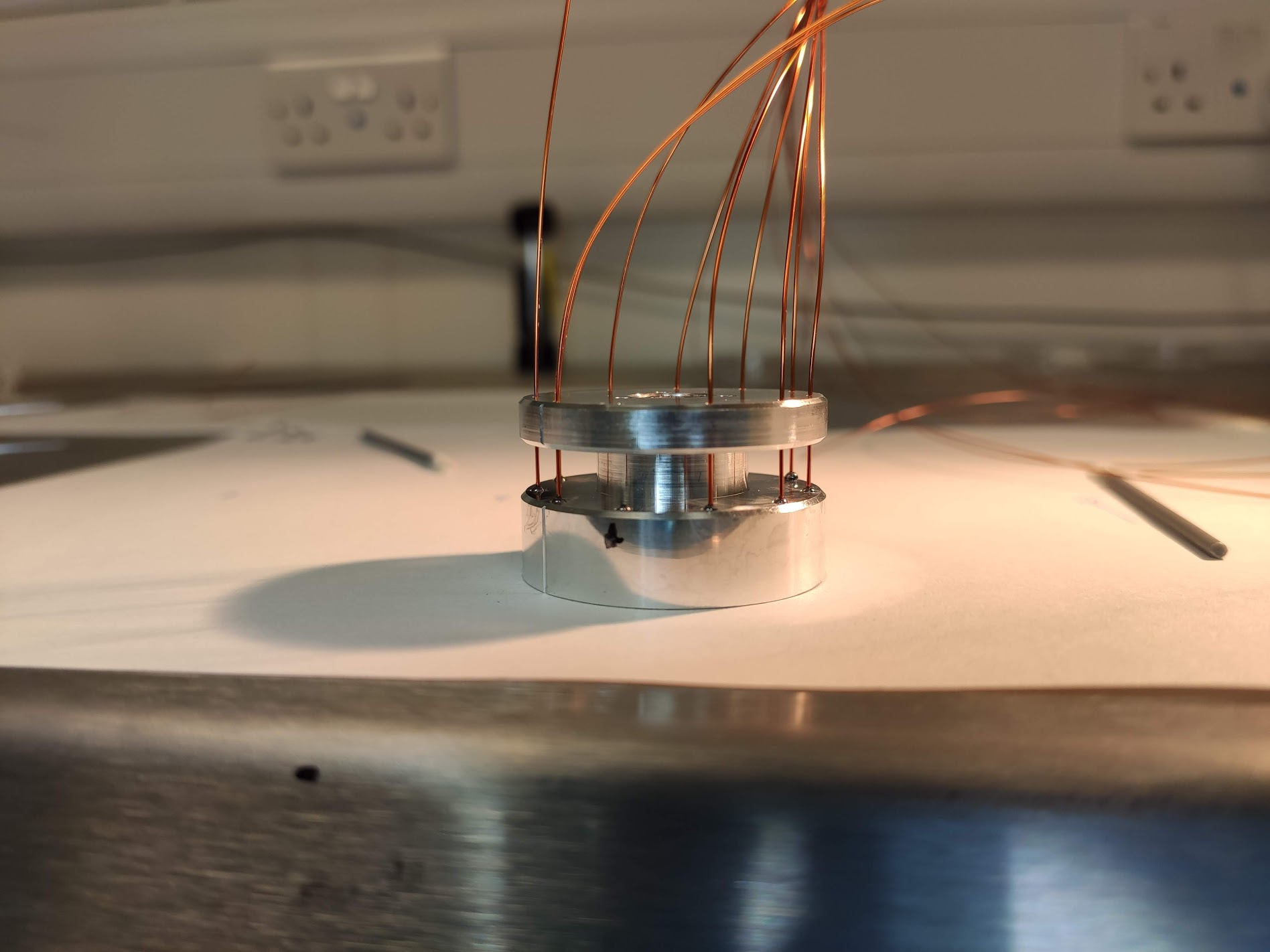}}
\subfigure[\label{fig:bhamAchinos}]{\includegraphics[width=0.25\textwidth]{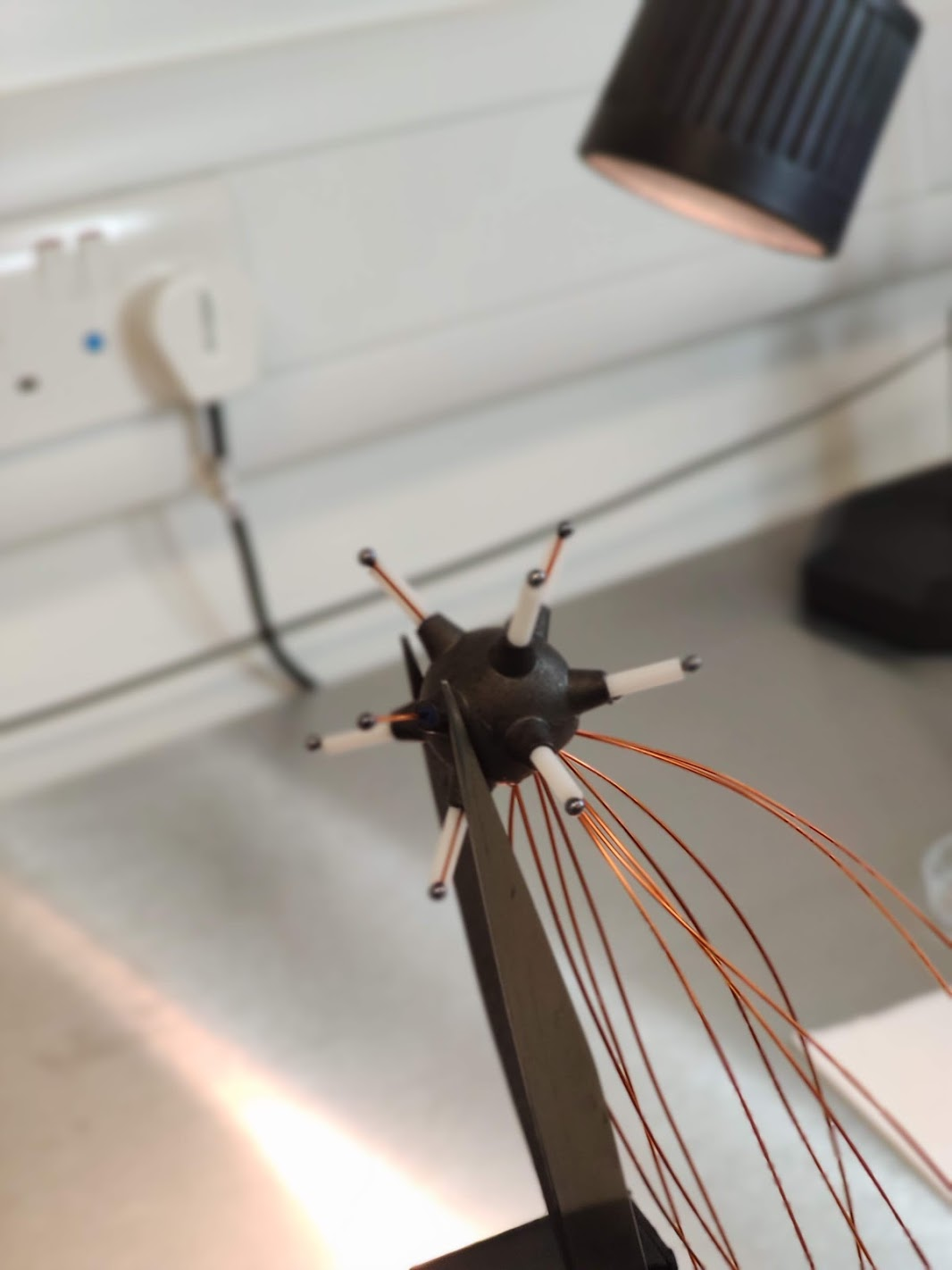}}
\caption{\subref{fig:bhamjig} Assembly tool used for simultaneously
  bonding several wires and anodes. \subref{fig:bhamAchinos} An
  ACHINOS being constructed using custom-made spacers to position and
  align the anodes.\label{fig:bhamLabAchinosandJig}}
\end{figure}

A major task is the industrialisation of the sensor manufacturing,
ensuring reliable and reproducible behaviour. The use of additive
manufacturing techniques, like 3D printing, for the construction of
the central electrode and its DLC coating has been a significant step
in this direction.
Further improvements in the manufacturing are
being explored, including the production of assembly tools for anode
alignment and attachment of wires. As an example, an assembly tool
constructed at the University of Birmingham, is shown in
figure~\ref{fig:bhamjig}, along with the spacers used to align the
anodes with during assembly, shown in figure~\ref{fig:bhamAchinos}.

To further improve the ACHINOS performance, the gain variation among
different anodes needs to be reduced. A major source of this variation
arises from irregularities in the shape of the
individual anodes. Identifying vendors that can supply anodes with
high sphericity and small surface roughness is being investigated.

A number of applications of the spherical proportional counter, such as
dark matter searches~\cite{Arnaud2018-nr}, Supernovae~\cite{Giomataris:2005fx} detection, fast neutron
spectroscopy~\cite{Savvidis2018-xo}, and measurements of neutrino
coherent scattering~\cite{Giomataris:2003bp}, would benefit from
operation under high gas pressure. The use of ACHINOS is an important
development in this aspect. To further progress in this direction, the
use of anodes with small radii, of order $100\;\si{\micro\meter}$, is
being explored. This would facilitate achievement of higher gain by
increasing the electric field near the anode surface, without further
increase of the anode voltage.  An important challenge towards smaller
anode radii is the bonding to the read-out wire, which needs to be
performed in a manner that minimises exposed contacts which may lead
to discharges. Ideas to achieve this are being explored with
industrial collaborators, at CERN and at CEA Saclay.

Furthermore, the number and configuration of anodes in the ACHINOS may
be optimised for each specific application, depending on the size and
pressure of the vessel. Such a developemnt can be guided by a
dedicated simulation program~\cite{Katsioulas:2020ycw}.  At the same
time, a further milestone would be the individual read-out and
high-voltage biasing of each anode. This would allow three dimensional
reconstruction of the ionisation track, and anode-by-anode calibration
of the gas gain.

\section{Summary}
A multi-anode read-out structure for the spherical proportional counter has been developed. 
\mbox{ACHINOS} now incorporates a DLC layer to prevent discharges, and its operation stability has been demonstrated. It was also found that the DLC layer is robust under discharges.
The gas gain in various pressures was also studied, demonstrating the ability to operate the detector in a gain of up to $\num[retain-unity-mantissa = false]{1E4}$ in $1500\;\si{\milli\bar}$ of Ar:CH${}_{4}$ ($98\%$:$2\%$).
The local energy resolution has been assessed as a function of azimuthal angle and found to be better than $10\%$ ($\sigma$). 
The difference in gain between different anodes was not greater than $25\%$, and this can be further improved by using anodes with a smaller deviation from an ideal sphere and lower surface roughness, individually calibrating each anode and the inclusion of more anodes.

\acknowledgments
This project has received funding from the European Union's Horizon 2020 research and innovation programme under the Marie Sk\l{}odowska-Curie grant agreement DarkSphere (grant agreement No~841261). Support has been received from the Royal Society International Exchanges Scheme. KN acknowledges support by the European Research Council (ERC) grant agreement no. 714893 and by UKRI-STFC through the University of Birmingham Particle Physics Consolidated Grant. We acknowledge the financial support of the Fundamental Research Funds for the Central Universities of China.

\bibliography{mybib}

\end{document}